\documentclass[12pt]{iopart}

\usepackage[T1]{fontenc}
\usepackage[ascii]{inputenc}
\usepackage{bm,cite}
\usepackage{graphicx}
\usepackage{color}
\usepackage{microtype}
\usepackage{iopams}  
\usepackage{hyperref}
\usepackage{upgreek}

\newcommand{\Real}{\mathrm{Re}\,}
\newcommand{\Imag}{\mathrm{Im}\,}
\newcommand{\text}[1]{\mathrm{#1}}
\newcommand{\ii}{\mathrm{i}}							
\newcommand{\ee}[1]{\mathrm{e}^{#1}}						
\renewcommand{\vec}[1]{\bm{#1}}							
\newcommand{\vectornorm}[1]{\left|\left| #1 \right|\right|}			

\newcommand{\braket}[2]{\left< #1 \vphantom{#2} \right|
 \left. #2 \vphantom{#1} \right>} 						
\newcommand{\ket}[1]{\left| #1 \right>}						
\newcommand{\del}[2]{\frac{\partial #1}{\partial #2}}				

\begin{document}

\title[Dipolar Bose-Einstein condensates in triple-well potentials]%
{Dipolar Bose-Einstein condensates in triple-well potentials}

\author{R\"udiger Fortanier, Damir Zajec, J\"org Main and
G\"unter Wunner}

\address{1. Institut f\"ur Theoretische Physik, Universit\"at Stuttgart,
70550 Stuttgart, Germany}
\eads{\mailto{ruediger.fortanier@itp1.uni-stuttgart.de},
\mailto{zajec@itp1.uni-stuttgart.de}}

\begin{abstract}
  Dipolar Bose-Einstein condensates in triple-well potentials are
  well-suited model systems for periodic optical potentials with
  important contributions of the non-local and anisotropic
  dipole-dipole interaction, which show a variety of effects such as
  self-organisation and formation of patterns. We address here a
  macroscopic sample of dipolar bosons in the mean-field limit. This
  work is based on the Gross-Pitaevskii description of dipolar
  condensates in triple-well potentials by Peter \etal 2012
  J.~Phys.~B:\ At.\ Mol.\ Opt.\ Phys.\ {\bf 45} 225302. Our analysis
  goes beyond the calculation of ground states presented there and
  clarifies the role of excited and metastable states in such systems.
  In particular, we find the formation of phases originating from the
  interplay of several states with distinct stability properties. As
  some of the phases are formed by metastable states special attention
  is paid to the characteristics of phase transitions in real-time and
  the dynamical stabilisation of the condensate.

\end{abstract}


\maketitle

\section{Introduction}
\label{sec:introduction}

One of the most interesting features of Bose-Einstein condensates
(BECs) is the possibility of the investigation of quantum effects in a
macroscopically controlled way. BECs have been experimentally realised
with atoms sustaining a large magnetic dipole moment such as
${}^{52}\mathrm{Cr}$ \cite{Griesmaier05a,Beaufils2008,Lahaye09} and,
more recently, $^{164}$Dy \cite{Lu2010,Lu2011_2} and
${}^{168}\mathrm{Er}$ \cite{Aikawa_2012}. Very recently, fast progress
towards the creation of BECs of polar molecules \cite{Ni_2008}, which
sustain large electric dipole moments, has been made. These
developments have opened the field of research of effects generated by
the dipole-dipole interaction (DDI). A major part of these effects can
be summarised under the topics ``self-organisation'' and pattern
formation. These effects are involved in the formation of a supersolid
quantum phase. Menotti \etal \cite{menotti07a} have shown by
calculations on the basis of the Bose-Hubbard model that the existence
of a supersolid is closely connected to the appearance of metastable
states in an optical lattice. However, the Bose-Hubbard description is
only practicable for small atom numbers and cannot describe the
collapse of the macroscopic wave function. The stability of these
metastable states with respect to the collapse thus has to be analysed
by the use of a method -- like the mean-field description -- which is
able to describe the local divergence of the wave function's
amplitude.

A minimal system for the analysis of the effects mentioned above is a
dipolar BEC in a triple-well (TW) potential. In Ref.~\cite{Lahaye10a}
the Bose-Hubbard model was used to investigate the possible ground
states for %
a mesoscopic sample of dipolar bosons up to 18 particles. There, four
different ground-state phases have been observed. Peter \etal
\cite{Peter12} applied a mean-field approach to this system, yet,
considering a macroscopic BEC. They found no clear separation of
phases, and observed that some population distributions predicted by
the Bose-Hubbard model are unstable within the Gross-Pitaevskii
description. Most of these calculations were performed on a grid using
imaginary-time evolution (ITE). While this is a globally convergent
method for the linear Schr\"odinger equation, the ITE does not
necessarily converge to the ground state in a nonlinear system such as
the GPE. In Ref.~\cite{menotti07a} this fact has been used to find
metastable states.

Dipolar BECs in double-well potentials have been investigated in
\cite{Xiong09a,Asad09}. In this system three major phases have been
found: In the first both wells are populated equally. The second is
the symmetry-broken phase (macroscopic quantum self-trapping, MQST),
in which the majority of the particles populates one well, and the
third is the unstable phase, where the condensate wave function
collapses. It has been pointed out in \cite{Asad09} and
\cite{Raghavan99} that MQST is a dynamical effect arising from the
interaction of the particles which reflects itself in the nonlinearity
of the GPE.

    In Ref.~\cite{Zhang12} Zhang \etal have performed a mean-field
    three-mode approximation to dipolar BEC in triple-well potentials.
    They found that the inter-level coupling of the states changes
    depending on their inter-site interactions and leads to
    macroscopic phase transitions. Furthermore, they show that the
    long-range nature of the dipole-dipole interaction leads to new
    dynamical effects such as long-range Josephson oscillations, where
    tunnelling between the outer wells takes place with negligible
    alteration of the population of the centre well. 

    Both approaches, the Bose-Hubbard model in Ref.~\cite{Lahaye10a}
    and the three-mode approximation in Ref.~\cite{Zhang12} cannot
    resolve the following answer which is crucial for a planned
    experiment: Are the states stable with respect to the collapse of
    the wave function when a macroscopic BEC is considered and
    therefore observable in an experiment?

\begin{figure}[tb]
  \centering
 \includegraphics[width=0.8\columnwidth]{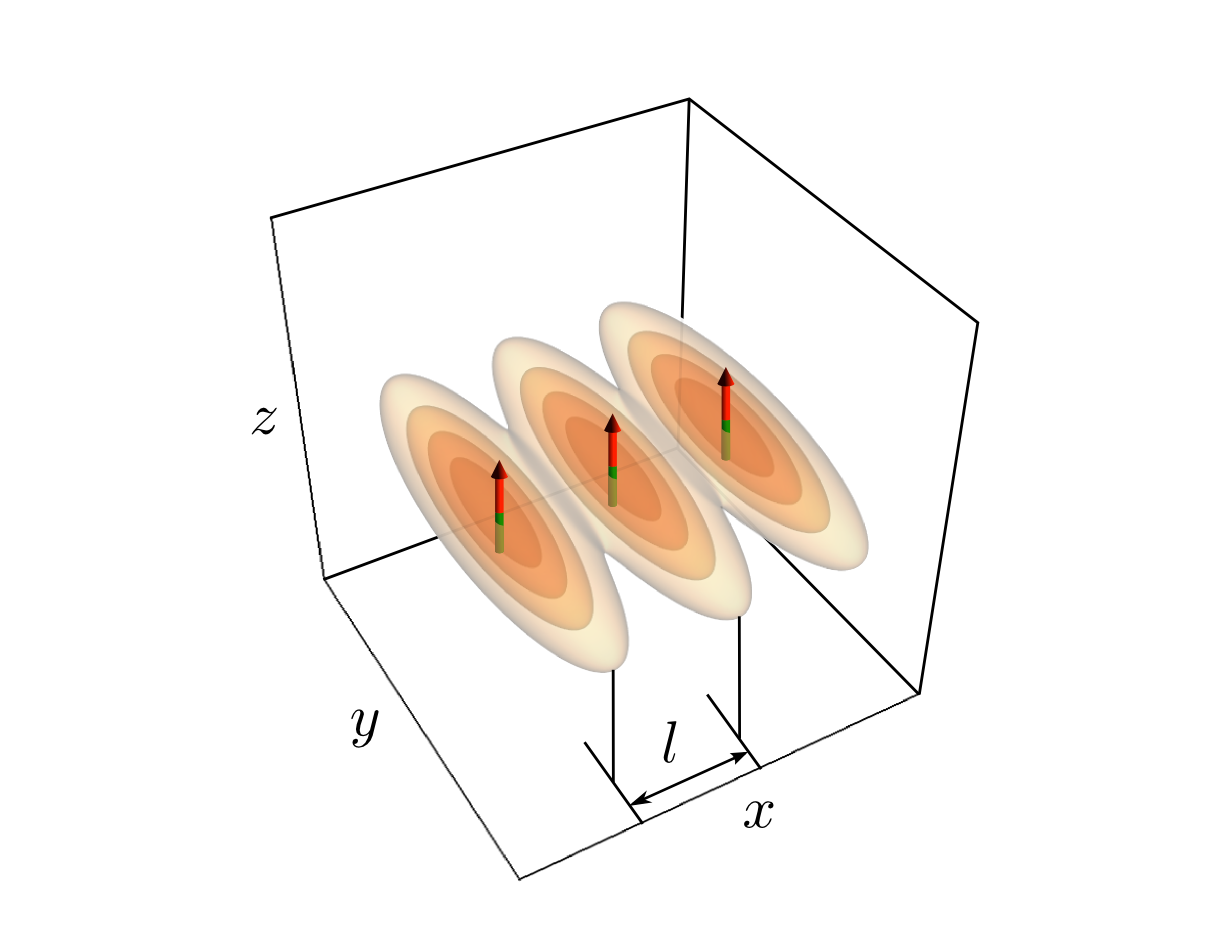}
 \caption{Visualisation of the TW potential. The
   dipoles are aligned in $z$-direction (repulsive configuration). The
   potential width $L_y=\omega_y/2$ is 4 times larger than $L_x$ and
   $L_z$. The parameter $l$ denotes the distance between the minima
   of the potentials.}
\label{fig:geometry}
\end{figure}

    This question can be answered by the use of the Gross-Pitaevskii
    equation as applied by Peter \etal \cite{Peter12}. There, however,
    no excited states were calculated which we find playing a crucial
    role in the formation of phases and the dynamical behaviour of the
    condensate. The purpose of this paper is to show that these
    states, which sometimes exist simultaneously, are the reason for
    quantum phase transitions and can be used to detect such by its
    dynamical characteristics. In particular, we investigate a dipolar
    BEC in an external TW potential on the basis of the extended
    time-dependent GPE in the form
\begin{eqnarray}
  \label{eq:GPE}
  H \Psi(\vec{r},t) &= \left( -\frac{1}{2}\Delta + V_{\mathrm{TW}} +
    V_{\mathrm{dd}} + V_{\mathrm{sc}} \right) \Psi (\bm{r},t)
  = \mathrm{i} \partial _{t}\Psi (\bm{r},t)\,,
\end{eqnarray}
with
\begin{eqnarray}
  V_{\mathrm{TW}} &= -V_0 \sum_{i=1}^3
  \text
{exp}\left(-\frac{2(x-q_x^i)^2}{\omega_x^2} 
  - \frac{2y^2}{\omega_y^2} - \frac{2z^2}{\omega_z^2}\right) \,,\nonumber \\
  V_{\mathrm{dd}} &= 3Na_{\mathrm{dd}} \int \text{d}^3r'
  \frac{1-3\cos^2\vartheta}{|\vec{r}-\vec{r}'|^3}
  |\Psi(\vec{r}',t)|^2 \,,\nonumber \\
  V_{\mathrm{sc}} &= 4\pi Na |\Psi(\vec{r},t)|^2 \,,\nonumber
\end{eqnarray}
where $N$ is the number of particles and $a_{\mathrm{dd}}$ and $a$
denote the dipole and scattering length, respectively. The centres of
the three individual wells are given by $q^1_x=-l$, $q^2_x=0$, and
$q^3_x=l$. The dipoles are aligned along the $z$-axis, so that
$\vartheta$ is the angle between the $z$-axis and the vector
$\vec{r}-\vec{r}'$. Here, we have adopted the unit system of Peter
\etal \cite{Peter12}, which implies measuring all lengths in units of
the inter-well spacing $l$, all energies in units of $\hbar^2/ml^2$
and time in units of $ml^2/\hbar$, where $m$ is the particle mass. The
TW potential and the orientation of the dipoles is visualised in
figure~\ref{fig:geometry} for the repulsive configuration (the
attractive configuration would imply the dipoles to be aligned in
$x$-direction). The widths of the Gaussians for the external trap are
chosen such that changing the polarisation direction does not change
the on-site effects of the dipole interaction $(\omega_x = \omega_z)$
and that the stability of the condensate is higher than in the
spherical case $(\omega_y > \omega_{x,z})$. Furthermore, $\omega_x$
has to be set to a value where the different wells are clearly
distinct. We further assume the relevant time scale of the inter-well
oscillations to be large in comparison with breathing-mode like
oscillations. This corresponds to a sufficiently low tunnelling rate.
The interesting effects mentioned above are expected to arise from the
interplay between the short-range and the long-range nature of the
interaction. We therefore included both kinds of interaction and
assumed $Na$ and $Na_\mathrm{dd}$ to be independently adjustable
quantities, whereas the results do not explicitly depend on $N$ due to
the scaling properties of the GPE.

The paper is organised as follows. In section~\ref{sec:methods} we will
briefly introduce the two methods we used to solve the GPE. In
section~\ref{sec:results} we will present our results and investigate the
system for two different sets of parameters in sections~\ref{sec:cut-06}
and \ref{sec:cut-02} in detail.

\section{Methods}
\label{sec:methods}

A well-known standard method to treat the dipolar GPE is the solution
on a grid, and we use this technique for the computation of ground
states and to simulate the real-time dynamics. However, the large
number of grid points (and therefore parameters) does not allow us to
obtain stationary states by a nonlinear root search. Thus, the
accessibility of excited states, which do play a crucial role in the
TW system, requires a larger effort, as it has been shown e.g.\ in
\cite{Rapedius2010a,menotti07a}. For this reason we also apply a
variational approach with coupled Gaussian wave packets (GWPs) which
has proven to be a full-fledged alternative to grid calculations
\cite{Rau10a,Rau10b,Eichler12a} for the description of the ground and
excited states as well as for real-time dynamics far beyond the
stationary solutions.

\subsection{Full-numerical grid calculations}
\label{sec:full-numerical-grid}

The propagation of the macroscopic wave function $\Psi$ in real time
and the ITE for the calculation of stationary solutions can be
performed on a grid. Particularly, ground states are calculated by the
evolution of an initial wave function in imaginary time $(t =
-\ii\tau)$ which dampens all other excited states.

For the linear Schr\"odinger equation the action of the time evolution
operator $U(\tau)$ on an initial state $\psi$ can be investigated by
expanding this state in the eigenfunctions $\phi$ of the Hamiltonian $H$
\begin{eqnarray}
  U(\tau)\ket{\psi} 
  = & \ee{-H\tau} \sum_{i} \ket{\phi_i} \braket{\phi_i}{\psi} = 
  \sum_{i} \ee{-E_i\tau} c_i \ket{\phi_i}.
\end{eqnarray}
Although the ITE dampens all of the eigenstates of the series, excited
states vanish faster than the ground state. This guarantees the global
convergence of the ITE. However, since in the nonlinear GPE the
Hamiltonian depends on the actual state, the basis set of the
expansion changes after each time step. Therefore the damping of the
excited states cannot be assured. We thus have to compare the
numerical results with the solution of the variational approach or
choose an initial wave function which is sufficiently similar to the
ground state. The latter can be realised by simply mapping the
solution of the variational approach on the grid or by using previous
numerical solutions with similar parameters. Even though we have not
computed excited states by means of the ITE, such states can still be
investigated on the grid when the solution of the variational approach
is used as an initial wave function for dynamical simulations of the
GPE.

We use the split-operator method for the grid calculations, where the
scattering potential $V_{\mathrm{sc}}$ and the dipole-dipole potential
$V_\mathrm{dd}$ have to be calculated at each time step. The latter
can be evaluated by means of the convolution theorem. Altogether, we
have to perform six Fourier transforms for each time step. A
comprehensive presentation of this approach is given in
\cite{Eichler12a}.

\subsection{Time-dependent variational approach}
\label{sec:TDVP-approach}

As an alternative to simulations on a grid the condensate wave
function can be parametrised by a set of variational parameters, and
the time evolution of the state is given by the time-dependence of the
variational parameters. Our variational ansatz consists of a linear
superposition of three GWPs. Each of the GWPs has the form
\begin{eqnarray}
  \label{eq:GWP-single-ansatz}
  g^k = \ee{-
      \left(
        \left(
          \vec x^T - \vec q^k
        \right)^T A^k
        \left(
          \vec x - \vec q^k
        \right) - \ii\left(\vec p^k\right)^T
        \left(
          \vec x - \vec q^k
        \right) +  \gamma^k
      \right)}\,,
\end{eqnarray}
where the symbol $T$ denotes the transposition and where in general
the time-dependent parameters $A^k$ are $3\times 3$ complex symmetric
matrices, $\vec p^k$ and $\vec q^k$ are real 3$d$ vectors, and
$\gamma^k$ are complex numbers. We assume that the $z$-direction (the
direction of the dipole alignment) has a strong confinement due to the
external trap and ignore translations and rotations in this direction
by setting $A^k_{xz}=A^k_{yz}=p^k_z=q^k_z=0$. However, for the other
directions we apply no further restrictions, particularly with respect
to position and movement of the GWPs in the $x$-direction. It is
reasonable to start with one GWP placed at the centre of each well.

To determine the time-development of the variational parameters  we
make use of the time-dependent variational principle (TDVP) in the
formulation of McLachlan
\cite{McLachlan1964a}
\begin{eqnarray}
  \label{eq:McLachlan}
  I = \vectornorm{\mathrm{i} \phi - H \Psi(t)}^{2}\stackrel{!}{=}\min\,,
\end{eqnarray}
where $\phi$ is varied and set $\phi \equiv \dot \Psi$ afterwards. The
variational wave function
\begin{eqnarray}
  \label{eq:ansatz_tdvp_eom}
    \Psi &= 
    \sum\limits_{k=1}^{3} g^k
\end{eqnarray}
is then inserted into \eref{eq:McLachlan} yielding the equations
of motion (EOM) for the variational parameters
\begin{eqnarray}
  \label{eq:eom}
  \dot{\vec z}^k = f\left( \vec z^k (t) \right) = f\left(A^k (t),\vec q^k (t),\vec
  p^k (t) ,\gamma^k (t) \right)\,.
\end{eqnarray}
Details can be found in \cite{Eichler12a}, where the same method has
been used to describe the collision of quasi-2$d$ anisotropic
solitons. Note that the method for the computation of the dipole
integrals $\langle V_{\mathrm {dd}}\rangle$ in (\ref{eq:GPE}) slightly
differs from the method used in \cite{Eichler12a}. In that work a
strong confinement of the external trap is assumed in the
$y$-direction perpendicular to the alignment of the dipoles, whereas
here a strong confinement is assumed in the $z$-direction parallel to
the alignment of the dipoles.

The stationary states are the fixed points of \eref{eq:eom} and
can be determined by a nonlinear root search (e.g.\ Newton-Raphson).
An alternative to find the real ground state is the application of ITE
to the EOM. However, as discussed in
section~\ref{sec:full-numerical-grid} the ITE does not always converge to
the ground state. In particular, if the initial wave function is close
to that of an excited state, the ITE will stay for a rather long
period in imaginary time on an plateau of almost the same mean-field
energy. In practice, it is not always possible to distinguish between
that case and the convergence to the ground state. To evolve the EOM
in imaginary time as well as in real time, a standard algorithm like
Runge-Kutta can be used. For a more detailed description see
Ref.~\cite{Eichler12a}.

The linear stability of the fixed points can be investigated by the
calculation of the eigenvalues $\Lambda=\Lambda^r+\ii \Lambda^i$ of
the Jacobian
\begin{eqnarray}
  \label{eq:jacobian}
  J= \del{\left( \Real \dot A^k, \Imag \dot A^k, \dot{\vec q^k}, \dot{\vec
        p^k}, \Real \dot{\gamma^k}, \Imag \dot{\gamma^k} \right)}{\left(
      \Real A^j, \Imag A^j, {\vec q^j}, {\vec p^j}, \Real
      {\gamma^j}, \Imag {\gamma^j} \right)}\,,
\end{eqnarray}
with $k,j=1 \dots 3$.
The eigenvalues appear in pairs of opposite sign
and correspond to excitations described by the
Bogoliubov-de~Gennes equations \cite{Kreibich13a,Rau10a,Rau10b}. If all
real parts $\Lambda^r = 0$, the fixed point is stable, otherwise it is
unstable.

\section{Results}
\label{sec:results}

Placing the wells in the attractive configuration (dipole alignment
$\rightarrow\rightarrow\rightarrow$) enforces the atoms to occupy the
centre well and does not show the same diversity of phases as the
repulsive configuration ($\uparrow\uparrow\uparrow$) does. We will
therefore focus on the repulsive configuration, which is visualised in
figure~\ref{fig:geometry} and offers the possibility to investigate
all effects of interest. For the TW potential $V_{\rm TW}$ in
(\ref{eq:GPE}) the identical parameters $V_0=80$ (this corresponds to
$\sim 16$ recoil energies \cite{Bloch2005}), $\omega_x= \omega_z =
1/2$, and $\omega_y = 4$ as given in \cite{Peter12} have been used.
\begin{figure}[]
  \centering
 \includegraphics[width=0.8\columnwidth]{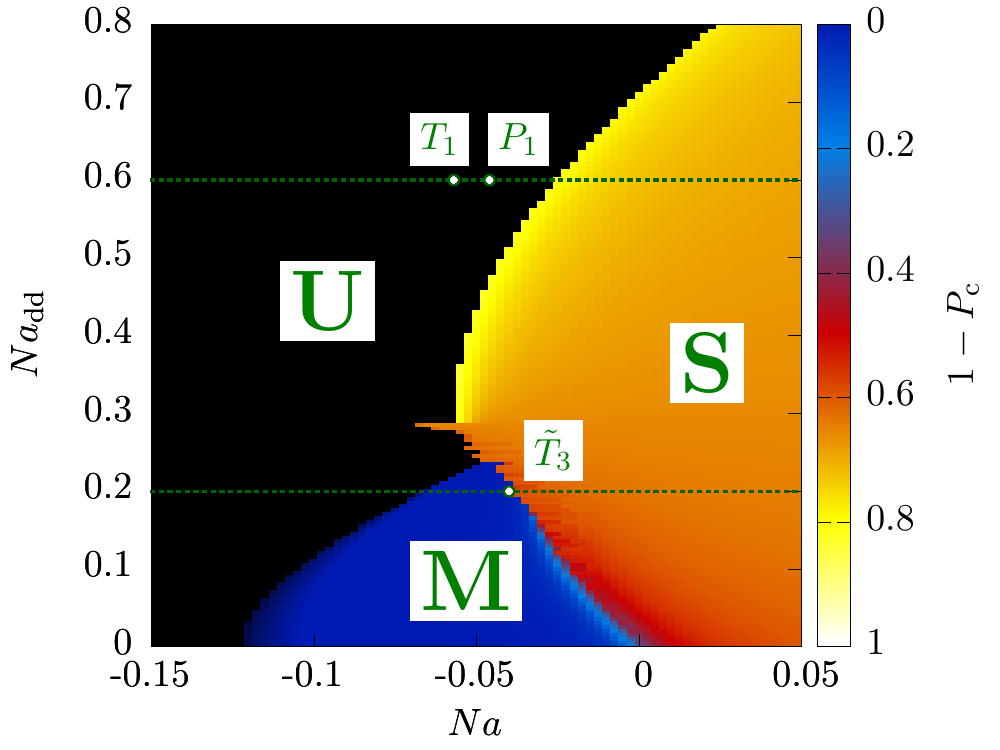}
 \caption{Phase diagram for the repulsive
   configuration. Coloured areas depict regions of parameter space
   where the ITE converges, whereas black areas indicate regions where
   the ITE does not converge. The colour bar represents the occupation
   of both outlying wells $1-P_\mathrm{c}$, where $P_\mathrm{c}$ is
   the population of the centre well. For details see the discussion
   in the text.}
\label{fig:results4_repulsive}
\end{figure} 
Figure~\ref{fig:results4_repulsive} shows the phase diagram for the
repulsive configuration obtained by grid calculations. 

Our aim here is to understand the nature of the distinct regions and
the mechanisms behind the phase transitions. The regions of the
parameter space where the ITE converges are depicted in grey-scale
(colour), whereas the black-coloured area $\mathrm U$ depicts the region
where no convergence occurs. The grey-scale (colour) bar on the
right-hand side shows the occupation of the two outlying wells with
$P_\mathrm{c}$ being the occupation of the inner well. This
means that the strip with $1-P_\mathrm{c}\approx 0.8$ for dipole
strengths $Na_{\text{dd}} \gtrsim 0.3$ represents states where most of
the particles are located in the outlying wells. Areas with
$1-P_\mathrm{c}\approx 2/3$ represent states where all three wells are
equally occupied.

The phase diagram shows an interesting feature for $Na_{\mathrm{dd}}
\lesssim 0.25$: At some critical value of the scaled scattering length
$Na$, the state, which the ITE converges to, exhibits a qualitative
sudden change. Here the number of time steps until a given criterion
for the convergence of the ITE is satisfied, reaches a local maximum.
Below this critical scattering length we find an area in the phase
diagram, marked as $\mathrm M$, where $1-P_\mathrm{c}\approx 0.1$
viz.\ almost all of the particles are located in the centre well.
Decreasing the scattering length even more finally leads to the
collapse of the condensate.

A similar result was presented by Peter \etal \cite{Peter12}. The
comparison with the phase diagram in \cite{Peter12}, which has been
calculated for the same set of parameters $Na$ and $Na_\mathrm{dd}$,
shows the following differences regarding the convergence of the ITE.
It does not show convergence of the ITE in the area $\mathrm M$,
whereas we do not find an additional white strip
($1-P_\mathrm{c}\approx 1$) which should be located right next to the
light (yellow) strip ($1-P_\mathrm{c}\approx 0.8$) for values of
$Na_{\text{dd}} \gtrsim 0.3$. This is not necessarily an
inconsistency, but indicates the existence of metastable states in the
system as will be discussed in detail in section~\ref{sec:cut-02}.

For a better understanding of the phase diagram in
figure~\ref{fig:results4_repulsive} and in particular to clarify the
nature of the phases U, S, and M and their transitions, we have
performed two different vertical cuts at $Na_\mathrm{dd}=0.6$ and
$Na_\mathrm{dd}=0.2$, marked by the dashed horizontal lines. Along
these lines, we calculated the stationary points by the use of the
variational ansatz enabling us to investigate ground and excited
states and their stability. In order to provide a deeper insight into
the dynamical properties beyond the linear vicinity of the stationary
states we subsequently performed real-time simulations.

\subsection{Cut at $Na_\mathrm{dd}=0.6$}
\label{sec:cut-06}

\begin{figure}[tb]
  \centering
 \includegraphics[width=0.7\columnwidth]{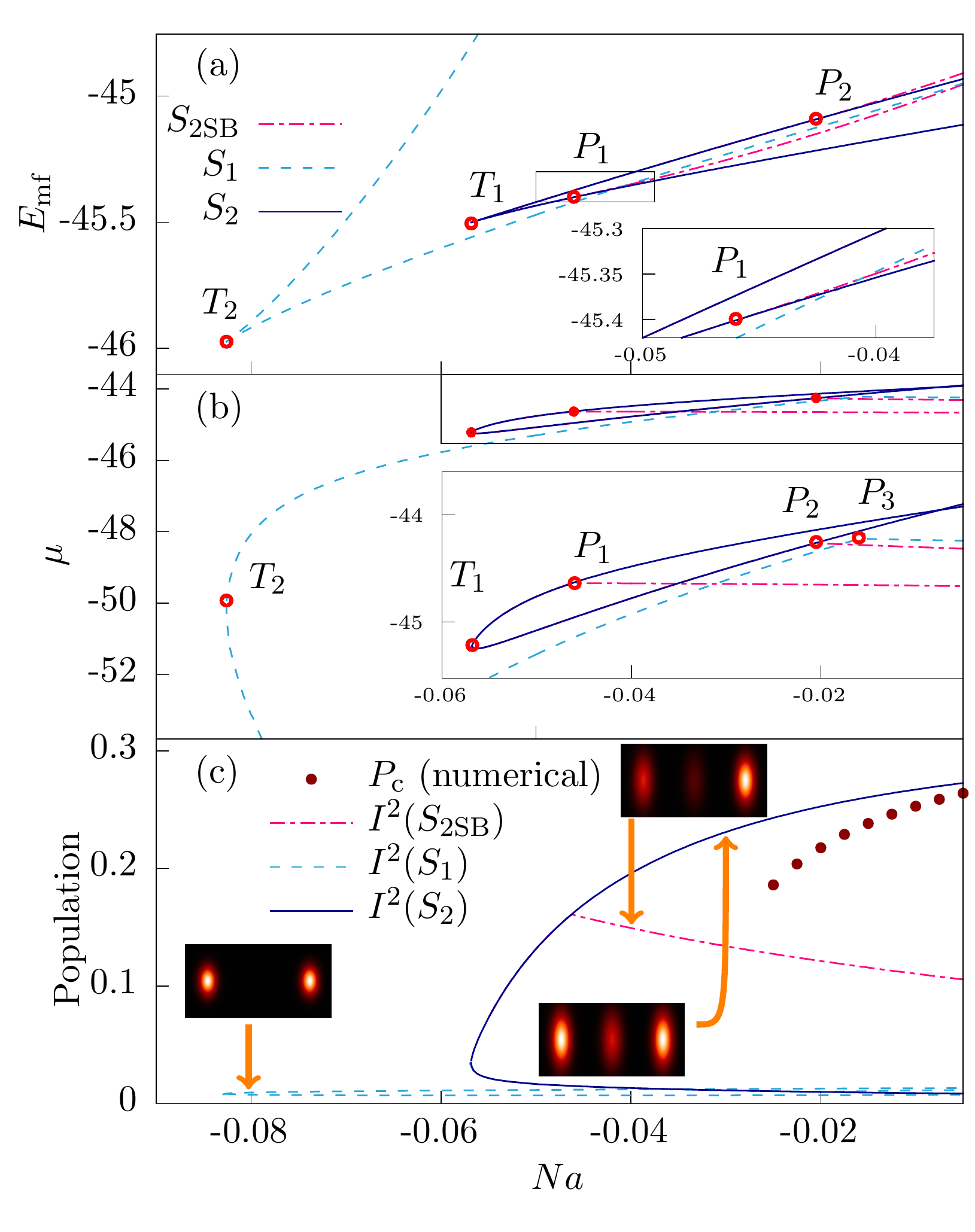}
 \caption{(a) Mean field energy and (b) chemical
   potential as functions of the scaled scattering length $Na$ for
   dipolar interaction $Na_\mathrm{dd}=0.6$. The insets show
   magnifications of the rectangles. (c) Overlap integral
   $I^{2}=\left< \left. g^2 \right| g^2 \right>$ which shows the
   population of the centre well. All states shown in (a) are
   represented. The dots show the results of the grid calculations
   obtained by the ITE. The identically scaled absorption images show
   the qualitative shape of the wave function at the positions, where
   the arrows are pointing to.}
\label{fig:variational-0-6}
\end{figure}

\begin{figure}[tb]
  \centering
 \includegraphics[width=0.7\columnwidth]{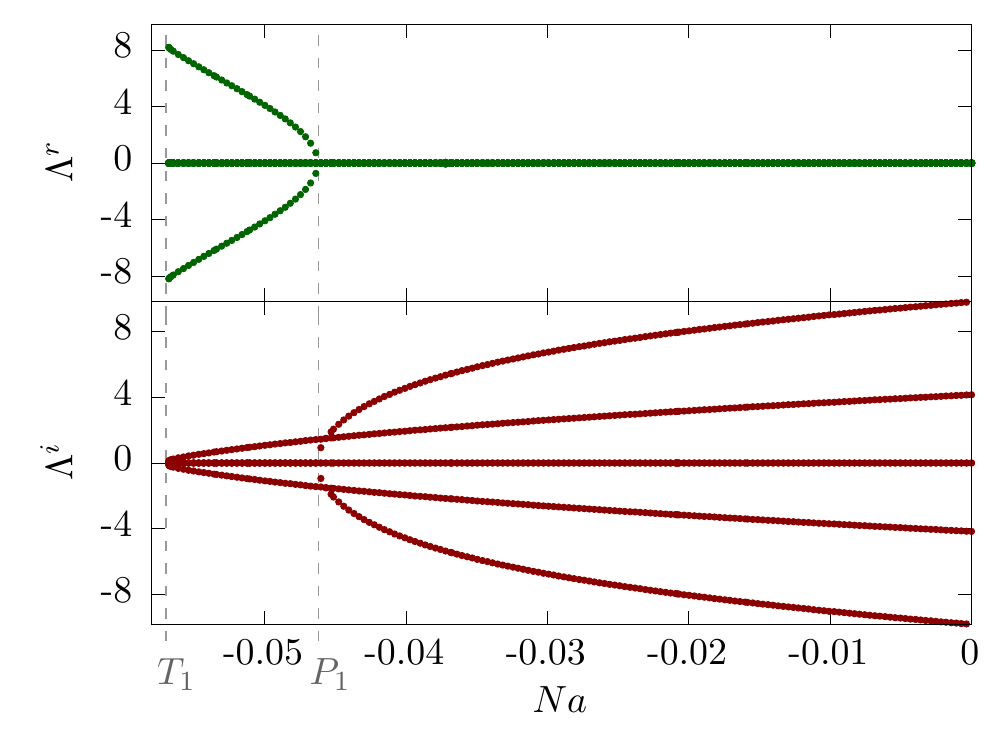}
 \caption{Eigenvalues $\Lambda$ of the Jacobian
   \eref{eq:jacobian} for the state $S_2$ with the lower mean-field
   energy (see figure~\ref{fig:variational-0-6}). The upper panel shows
   the real parts, the lower panel shows the imaginary parts. The
   state emerges unstable at the tangent bifurcation $T_1$ which can
   be seen from the non-vanishing $\Lambda^r$. At the pitchfork
   bifurcation $P_1$ the state becomes stable and can be viewed as the
   stable ground state from there on.}
\label{fig:eigenvalues}
\end{figure}

The results for $Na_{\mathrm{dd}}=0.6$ are shown in
figures~\ref{fig:variational-0-6} and \ref{fig:eigenvalues}. In
figure~\ref{fig:variational-0-6} we immediately see that more than one
state can be found. There are two tangent bifurcations $T_1$ and $T_2$
where two states emerge, respectively. We will denote these two pairs
of states with $S_1$ and $S_2$, and the symmetry-broken states
bifurcating from $S_2$ (see below) with $S_{2\mathrm{SB}}$. If one
investigates the linear stability by calculating the eigenvalues of
the Jacobian in \eref{eq:jacobian} one finds that at the tangent
bifurcations all four born states are unstable. For the states born at
$T_2$ this stays true for all values of $Na$, but not for the states
emerging at $T_1$.

In figure~\ref{fig:eigenvalues} the eigenvalues of the Jacobian are
plotted for the $S_2$-state with the lower mean-field energy. It
passes through a pitchfork bifurcation at $P_1$ and becomes stable for
higher values of $Na$. It can be seen as the stable ground state from
there on. The point $P_1$ is shifted to lower values compared to the
border obtained by numerical grid calculations. However, the exact
position of this border depends on the numerical method (e.g.\ the
choice of initial conditions for the ITE; cf.~\cite{Peter12}).
Furthermore, the quantitative results of the variational solution is
limited by the restrictions of the ansatz which uses only one GWP per
well. We expect the variational results to converge with increasing
number of GWPs \cite{Rau10b}. In the bifurcation $P_1$ two more states
$S_\mathrm{2SB}$ are involved. The energies of these states are
degenerate and the wave function breaks the symmetry of the trap (for
one of them the left well is populated more than the right well, for
the other one v.v.). The second state born in $T_1$ also passes
through a pitchfork bifurcation at $P_2$ but becomes stable only in
one of the eigenvalues while other unstable directions exist.

In figure~\ref{fig:variational-0-6}c we plot the quantity $I^k =
\braket{g^k}{g^k}$ which is a good estimate for the population of the
$k^{\mathrm{th}}$ well, if we assume a small overlap of the GWPs. The
states emerging at $T_2$ show that the middle well is hardly
populated. Nearly all particles are in the outer wells. We will call
this a split-state. This is not the case for both of the states
emerging at $T_1$, where (for the state with the lower mean-field
energy) some particles are in the middle well. Note that the
split-state is passing through such a symmetry-breaking pitchfork
bifurcation at $P_3$ as well. In fact, all states we investigated show
this kind of symmetry-breaking behaviour, yet we did not analyse those
in all cases.

An intriguing feature is the occurrence of a region between $P_1$ and
the crossing of the mean-field energy of the lower state of $S_1$ and
$S_2$, where we find an unstable state with a lower energy than the
stable ground state (see inset in figure~\ref{fig:variational-0-6}a).
This is only possible due to the nonlinearity of the GPE.
Nevertheless, it is interesting to perform real- and imaginary-time
calculations in this region. It turns out that the ITE converges for a
wide range of initial conditions to the stable ground state. However,
if one starts close to the unstable state one finds a large plateau in
the ITE, which is an indicator for the existence of metastable states.

The real-time evolution of the unstable states reveals that all of
them have a small life time and mostly end in the collapse of the wave
function. Calculations in which we decreased the scattering length as
a function of time from the stable ground state to values of $Na$
below $P_1$ showed a similar behaviour. The dynamical properties become
more complex at $Na_{\mathrm{dd}}=0.2$ and will be discussed in more
detail in section~\ref{sec:cut-02} where we will find a different
scenario.

The results we have found provide a better understanding of the
ITE-behaviour. In the calculation of the phase diagram in
figure~\ref{fig:results4_repulsive} for each $Na$ the result of the ITE
at the previous value of $Na$ has been taken as the initial wave
function. Obviously, this procedure gives rise to difficulties if
crossings of states and metastable states are involved due to the
small damping of states with almost the same energy as the ground
state.

\subsection{Cut at $Na_\mathrm{dd}=0.2$}
\label{sec:cut-02}

\begin{figure}[tb]
  \centering
 \includegraphics[width=0.8\columnwidth]{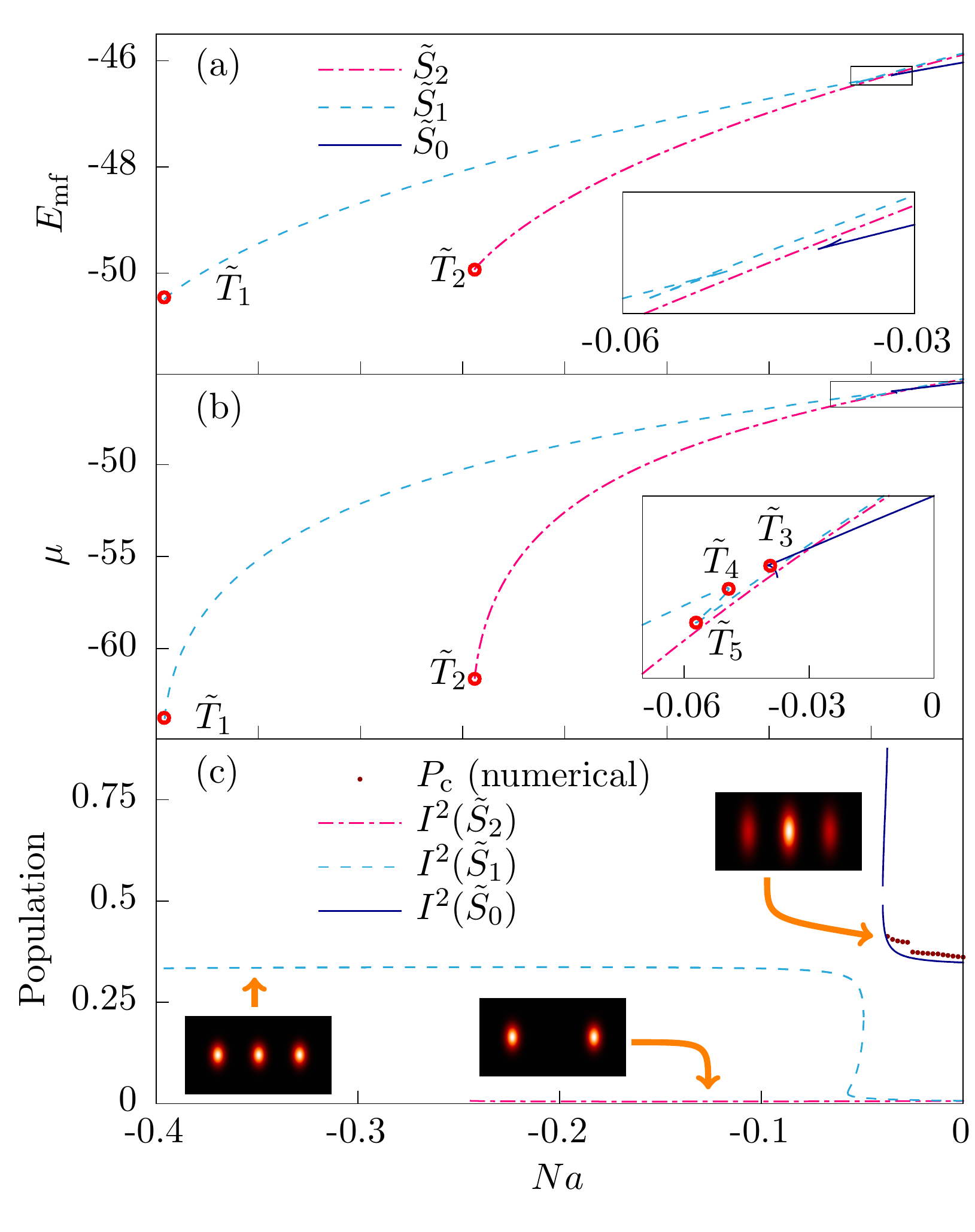}
 \caption{(a) Mean-field energy and (b) chemical
   potential as functions of the scaled scattering length $Na$ for
   dipolar interaction $Na_\mathrm{dd}=0.2$. (c) Overlap integral
   $I^{2}$ (c.f. figure~\ref{fig:variational-0-6}) which shows the
   population of the centre well. The absorption images show the
   qualitative shape of the wave function at the positions, where the
   arrows are pointing to.}
\label{fig:variational-0-2}
\end{figure}

We performed the second cut at $Na_{\mathrm{dd}} = 0.2$. The results
for the mean-field energy, chemical potential and the population of
the centre well are shown in figure~\ref{fig:variational-0-2}a-c,
respectively.
There are essentially three states $\tilde S_1$, $\tilde S_2$, and
$\tilde S_0$ emerging in the tangent bifurcations $\tilde T_1$,
$\tilde T_2$, and $\tilde T_3$, respectively. The $\tilde S_0$-state
with the lower mean-field energy born at $\tilde T_3$ is the stable
ground state. It is stable for the whole range of $Na$. The critical
point where it becomes unstable therefore is given by the tangent
bifurcation $\tilde T_3$ (and not by any stability change in a
pitchfork bifurcation, as for $Na_{\mathrm{dd}}=0.6$ in
figure~\ref{fig:variational-0-6}). Furthermore, the energy of the ground
state stays the lowest one for all $Na$, and no other state is
crossing. The excited states emerging at $\tilde T_1$ and $\tilde T_2$
have been omitted for reasons of clarity. While $\tilde S_2$ is a
split-state for all $Na$, $\tilde S_1$ shows some interesting behaviour
when it becomes energetically close to the ground state. It passes
through two consecutive tangent bifurcations $\tilde T_4$ and $\tilde
T_5$. It can be seen in figure~\ref{fig:variational-0-2}c that this
involves a qualitative change of the wave function's nature from a
state, where all wells are populated equally to a split-state.

\begin{figure}[tb]
  \centering
 \includegraphics[width=0.8\columnwidth]{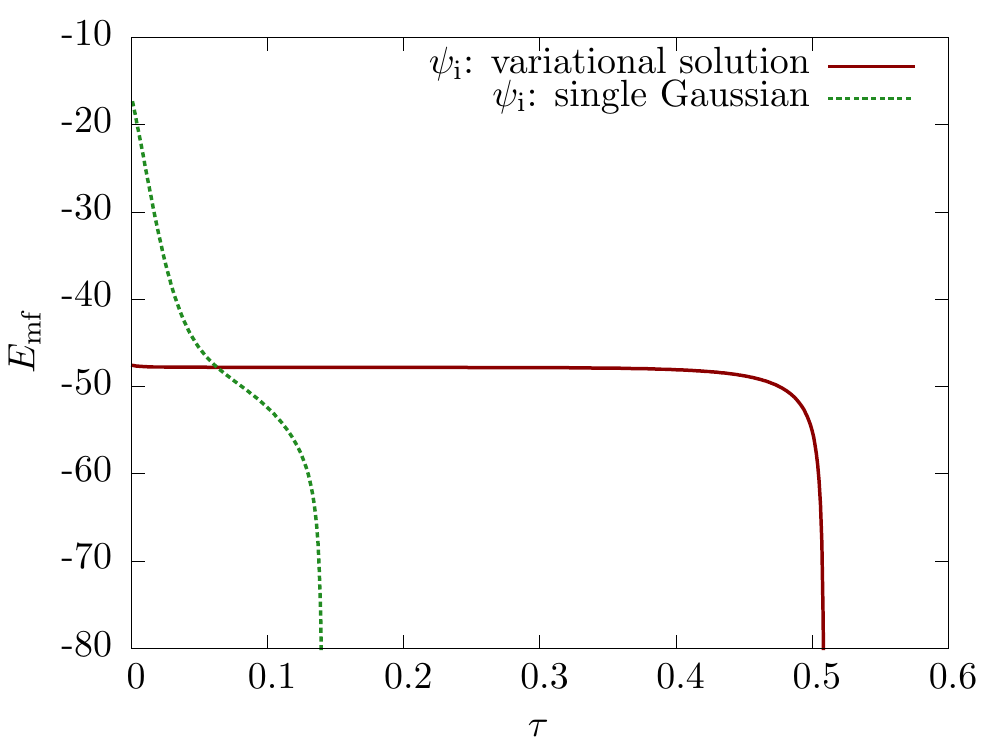}
 \caption{Mean-field energies as functions of imaginary
   time $\tau$ for two grid-ITEs at $Na_\mathrm{dd}=0.2$, $Na=-0.2$,
   where we have used as initial wave functions $\psi_\mathrm{i}$ a
   single Gaussian covering all wells for the calculation plotted as
   green dashed line and the variational solution of the state $\tilde
   S_2$ plotted as red solid line.}
\label{fig:vergleich_var_gauss}
\end{figure}

It is a remarkable fact that we find (unstable) states for values of
$Na$ where in figure~\ref{fig:results4_repulsive} no ground state could
be found, at all. More precisely, these states exist far below the
critical scattering length at the tangent bifurcation $\tilde T_3$.
This suggests a relation to the occurrence of regions in the parameter
space such as the area $\mathrm M$. The variational calculations
predict that no stable ground state is present in the area $\mathrm
M$. Although the grid-ITE seems to converge in this area, this could
still be just an effect of metastable states being present. This is
illustrated in figure~\ref{fig:vergleich_var_gauss}, where grid-ITEs
with two different initial states $\Psi_\mathrm{i}$ are shown for
$Na=-0.2$, which is in an area where no stable ground state exists.
The closer the initial state to the metastable state is, the more
pronounced the plateau in the ITE becomes. Finally, both calculations
diverge, indicating that the state is unstable.

For a deeper insight into the physics of the metastable states we have
investigated the dynamical properties in these regions. Real-time
evolutions in which we change the scaled scattering length $Na$ over
time show that the dynamics of the condensate in this region of the
parameter space differs from the dynamics in other regions.

\begin{figure}[]
  \centering
 \includegraphics[width=0.8\columnwidth]{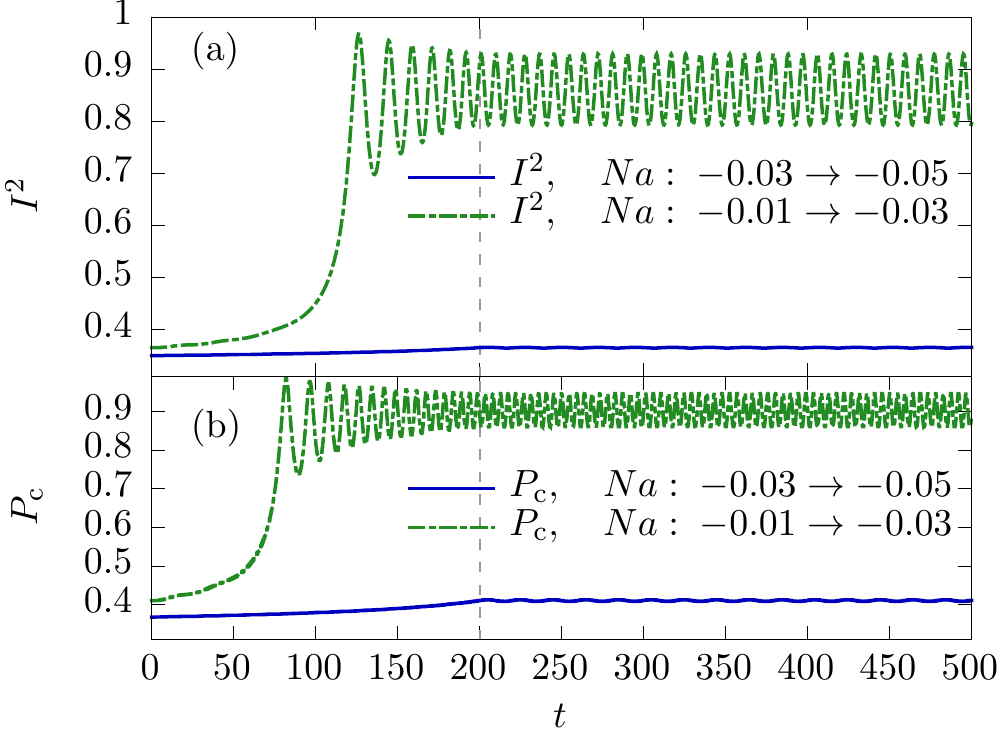}
 \caption{Real-time evolution for two transitions of
   $Na$. The scaled scattering length $Na$ is tuned linearly from
   $t=0$ to its final value at $t=200$ and kept constant from there
   on. (a) Variational calculation with the overlap integral
   (introduced in figure~\ref{fig:variational-0-6}c) of the middle well.
   (b) Population of the centre well obtained by the grid calculations
   is plotted as a function of time. In both panels (a) and (b) the
   solid blue line belongs to the calculation, where $Na$ is tuned
   from $-0.03$ to $-0.05$ and the dashed green line belongs to the
   transition from $Na=-0.01$ to $Na=-0.03$.}
\label{fig:real_time_evolution_grid}
\end{figure}

Furthermore, the dynamical tuning of the scattering length can be a
procedure in an experiment to access the metastable region. We
illustrate this in figure~\ref{fig:real_time_evolution_grid}a and b
for the variational and grid calculations, respectively, where we have
calculated the occupation of the centre well for two different
real-time evolutions. The solid blue and dashed green curve represent
the real-time evolutions for $Na_{\text{dd}}=0.2$ and an alteration of
$Na$ from $-0.01$ to $-0.03$ and from $-0.03$ to $-0.05$,
respectively.
For the solid (blue) curve, the alteration of the scattering strength
$Na$ starts and ends in the region S (see
figure~\ref{fig:results4_repulsive}) of the parameter space. This
leads to small oscillations in the occupation of the centre well as
soon as $Na$ is adjusted to its final value at $t=200$. For the dashed
(green) curve small oscillations are visible as soon as we change the
scattering length. Here, the alteration of $Na$ ends in regions of the
parameter space where the phase diagram given in \cite{Peter12} and
the one presented in figure~\ref{fig:results4_repulsive} differ. In
this case, during the ramp-down of the scattering length the
qualitative shape of the wave function changes significantly and large
periodic oscillations set in. This indicates the phase transition from
region S to M in figure~\ref{fig:results4_repulsive}. We conclude from
these calculations that the metastable states can prevent the
condensate from the collapse in some cases where no stable ground
state exists.

In an experiment absorption imaging of the condensate during the
ramp-down of the scattering length could reveal the phase transition.
For an experimental setup we adopt the experimental parameters of
Ref.~\cite{Lahaye10a} where a spacing of $l = 1.7 ~\upmu$m is
suggested. This setup can be realised by the use of $^{52}$Cr atoms
where $a_{\mathrm{dd}} \sim 0.79$ nm and $a \sim 5.8$ nm. The
presented transition in parameter space would be equal to a number of
420 $^{52}$Cr atoms where the scattering length would be tuned over a
time of 475 ms via Feshbach resonances \cite{Lahaye2007} to its final
value. This alteration of the scattering length though can be
performed much faster, albeit with the consequence of enhanced
oscillations in the centre well due to stronger excitations caused by
the faster tuning of the scattering length.

\begin{figure}[]
  \centering
 \includegraphics[width=0.8\columnwidth]{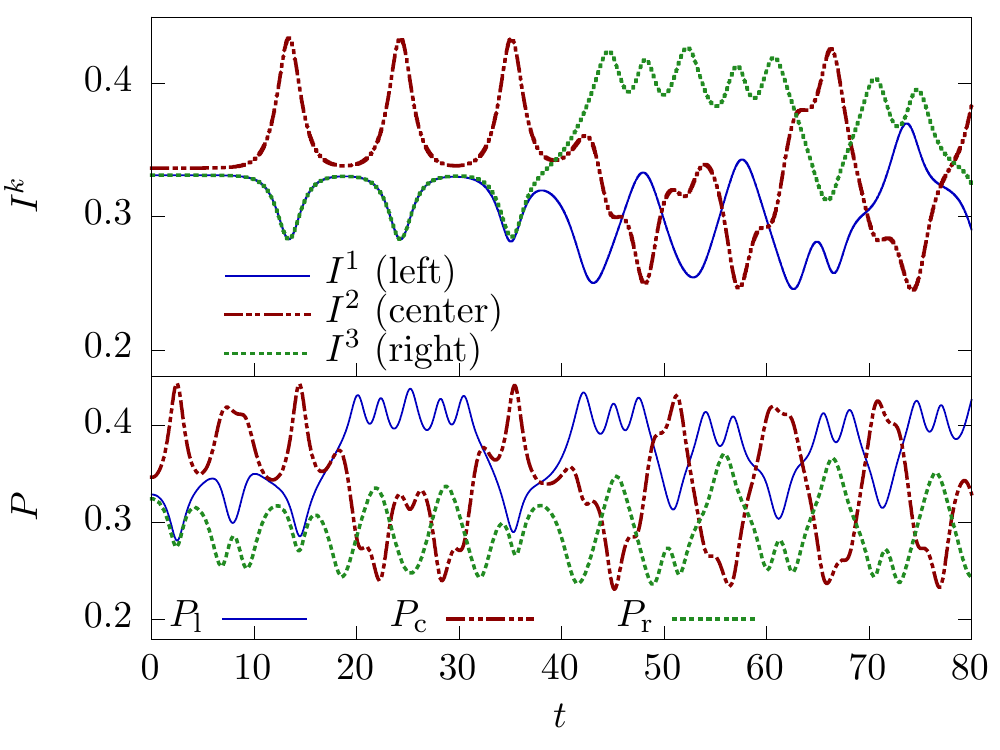}
 \caption{Real-time evolution of a metastable state for
   $Na_{\mathrm{dd}}=0.2$ and $Na=-0.2$. The variational approach has
   been used for the calculation shown on the upper panel. The figure
   shows the overlap integrals $I^{k}$ as already defined in
   figure~\ref{fig:variational-0-6}. The lower panel shows the results
   of the grid calculation for the population $P$ of the left, centre,
   and right well, respectively.}
\label{fig:rte-add0-2}
\end{figure}

If we use a single Gaussian function for the initial wave function of
the ITE (dashed green line in figure~\ref{fig:vergleich_var_gauss}), we
observe a quick divergence of the wave function in imaginary time.
Moreover, the real-time evolution of all states passed during this ITE
yields an almost instant collapse of the condensate. However, if we
choose a wave function from the plateau of the solid red line in
figure~\ref{fig:vergleich_var_gauss} where the slope is minimal, we
obtain a state where the real-time evolution does not yield an instant
collapse, as it can be seen in figure~\ref{fig:rte-add0-2}. The
variational calculation reveals that the corresponding state is
metastable. From figure~\ref{fig:variational-0-2} we can see that no
stable ground state exists for the set of parameters. The initial
state has been calculated with high precision by the nonlinear root
search. We therefore observe that the population of the wells stays
constant for some time (see upper panel of figure~\ref{fig:rte-add0-2}).
Then, a quasi-periodic symmetric oscillation sets in, where the
population of the outer wells and the inner well is modulated
periodically. After several oscillations the symmetry of this unstable
oscillation is broken and chaotic oscillations between the wells set
in.

In the grid calculations (see lower panel of
figure~\ref{fig:rte-add0-2}) the population stays constant for a shorter
time. This indicates that the metastable state is not perfectly hit
(which is also visible by the slight symmetry-breaking of the initial
state). Consequently, a complete break of the symmetry is reached
earlier. The chaotic oscillations have a slightly higher frequency in
the grid calculations. This is a consequence of the variational ansatz
reducing the tunnelling between the wells.

The dynamics shown in figure~\ref{fig:rte-add0-2} indicate that the
unstable states contribute to the dynamical behaviour of the metastable
states. In fact, we qualitatively find configurations where all wells
are populated equally, split states, and symmetry-broken states during
the real-time evolution of this metastable state.

\section{Conclusion}
\label{sec:conclusion}

We studied the ground and metastable states of dipolar BECs in
triple-well potentials both with a full-numerical ansatz and a
time-dependent variational principle with coupled Gaussians. Although
the triple-well potential constitutes a very simple system, the
occurrence of different phases and their stability properties appears
to be quite complicated. The phase diagram presented depicts a region
of the parameter space, where multiple phases occur. This includes
states where all wells are equally occupied as well as states where
more particles are located in the centre or the both outer wells. The
variational solutions reveal a variety of excited and unstable states
including symmetry-broken states and unstable states in regions where
no stable ground state exists. Real-time evolutions and the
eigenvalues of the Jacobian $J$, given in \eref{eq:jacobian}, both
allow for predictions about the stability of the investigated states.
Moreover, the former is able to show whether the occupation of the
wells for metastable states is characterised by quasi-periodic,
chaotic oscillations, a break of symmetry, or whether the instability
leads to a collapse. We have pointed out that a dynamical
stabilisation of the condensate by the interplay of the metastable
states is possible in those regions. Therefore, these regions are best
candidates for the observation of a supersolid phase.

Further investigations should include multi-well potentials with
additional wells and a different arrangement like a triangular or
ring-like configuration. Our results should stimulate experimental
efforts to study dipolar BECs in multi-well potentials.

\section*{Acknowledgements}
We thank David Peter and Tilman Pfau for valuable discussions. R.E.\
is grateful for support from the Landes\-graduierten\-f\"orderung of
the Land Baden-W\"urttemberg. This work was supported by Deutsche
For\-schungs\-ge\-mein\-schaft.

\section*{References}

\providecommand{\newblock}{}


\end{document}